\documentclass{aa}  
\usepackage{graphicx}
\usepackage{txfonts}
\newcommand{\Ms}{$\textrm{M}_{\odot}$}
\newcommand{\SFR}{\textrm{M}$_{\odot}$~\textrm{yr}$^{-1}$}
\newcommand{\kms}{$\textrm{km~s$^{-1}$}$}
\newcommand{\Ha}{H$\alpha$} 

\begin{document}

   \title{Star formation in outer rings of S0 galaxies. II.}

   \subtitle{NGC 4513 -- a multi-spin ringed S0 galaxy.}

   \author{I. Proshina
          \inst{1}
          \and
          O. Sil'chenko
          \inst{1}
          \and
          A. Moiseev
          \inst{2,1}
          }

   \institute{Sternberg Astronomical Institute of the Lomonosov Moscow
             State University, University av. 13, 119234 Russia\\
             \email{ii.pro@mail.ru,olga@sai.msu.su}
          \and
             Special Astrophysical Observatory
             of the Russian Academy of Sciences,
             Nizhnij Arkhyz, 369167 Russia\\
              \email{moisav@gmail.com}
             }

   \date{Received  .., 2019; accepted .., 2019}

 
  \abstract
   {}
   {Though S0 galaxies are usually thought to be `red and dead', they
   demonstrate often star formation organized in ring structures. We try
   to clarify the nature of this phenomenon and its difference from star
   formation in spiral galaxies. The moderate-luminosity nearby S0 galaxy, NGC 4513, is studied here.}
   {By applying long-slit spectroscopy along the major axis of NGC 4513, we have measured gas and star
kinematics, Lick indices for the main body of the galaxy, and strong emission-line flux ratios in the ring. 
After inspecting the gas excitation in the ring using the line ratios diagnostic diagrams 
and have assured that it is ionized by young stars, we have determined the gas oxygen abundance by
using popular strong-line calibration methods. We have estimated star formation rate (SFR) in the outer
ring by using the archival Galaxy Evolution Explorer (GALEX) ultraviolet images of the galaxy.}
   {The ionized gas counterrotates the stars over the whole extension of NGC 4513 so being accreted from outside.
The gas metallicity in the ring is slightly subsolar, [O/H]=-0.2 dex, matching the metallicity of the stellar
component of the main galactic disc. However the stellar component of the ring is much more massive
than can be explained by the current star formation level in the ring. We conclude that probably the ring of NGC~4513
is a result of tidal disruption of a massive gas-rich satellite, or it may be a consequence of a long star-formation event
provoked by a gas accretion from a cosmological filament having started some 3~Gyr ago.}
   {}


   \maketitle
%

\section{Introduction}
Outer rings are common attributes of S0 galaxies by their definition \citep{devauc59}.
The imaging statistics reveals that about 50\%\ S0-S0/a galaxies possess outer stellar
rings \citep{arrakis,nirs0s}. Among those, about 50\%\ are also seen in the UV-bands \citep{kostuk15}
so experiencing recent star formation and containing probably some amount of gas to fuel
this star formation. The { cool} gas origin in S0s is still vague:
{ though it is present in the most S0 galaxies \citep{welchsage03, welchsage06,welchsage10}, but its spin is often decoupled from that of the stellar 
component \citep{bertola92,kuijken96,kf2001,padua2004,atlas3d_10,atlas3d_13}, 
especially in rarefied environments \citep*{isomnras,isosalt}, that gives the evidence for recent gas accretion from outside \citep{thakar_ryden96,thakar_ryden98} along an arbitrary direction. Even much less} is known about star formation { in S0s
providing {\it stellar} ring structures: it proceeds only in about the half of gas-rich lenticular galaxies \citep{pogge_esk93}, and the conditions provoking star formation occurrence in the gas accreted by S0s are not completely understood.} 

\begin{table} 
\caption[ ] {Global parameters of the galaxy} 
\begin{flushleft}
\begin{tabular}{lc} 
\hline
\noalign{\smallskip}
Galaxy & NGC 4513  \\ 
Type (NED$^1$) & (R)SA0$^0$\\ 
$R_{25}$, (NED$+$RC3$^2$) & 43\arcsec\ or 7~kpc\\ 
$B_T^0$ (LEDA$^3$) &  13.88 \\ 
$M_B$ (LEDA)  & --18.99  \\ 
$M_H$ (NED)  & --22.84  \\ 
$V_r $ (NED) & 2304 $\mbox{km} \cdot \mbox{s}^{-1}$\\ 
Distance, Mpc (NED)  & 33 \\ 
Inclination (LEDA) & $59^{\circ}$  \\
{\it PA}$_{phot}$ (LEDA)  &  $15.7^{\circ}$ \\ 
$V_{rot} \sin i$, $\mbox{km} \cdot\mbox{s}^{-1}$, (HI$^4$) & $\sim 170$ \\ 
$\sigma _*$, $\mbox{km} \cdot\mbox{s}^{-1}$, (LEDA) & 120 \\ 
$M_{HI} ^4$, $10^9\,M_{\odot}$ & 0.27 \\
\hline
\multicolumn{2}{l}{$^1$\rule{0pt}{11pt}\footnotesize NASA/IPAC Extragalactic
Database, http://ned.ipac.caltech.edu .}\\ 
\multicolumn{2}{l}{$^2$\rule{0pt}{11pt}\footnotesize Third
Reference Catalogue of Bright Galaxies, \citet{rc3} .}\\
\multicolumn{2}{l}{$^3$\rule{0pt}{11pt}\footnotesize Lyon-Meudon Extragalactic
Database, http://leda.univ-lyon1.fr .}\\ 
\multicolumn{2}{l}{$^4$\rule{0pt}{11pt}\footnotesize
\citet{tang_hi}}\\
\end{tabular} 
\end{flushleft} 
\end{table}

\begin{figure*}[htb!]
   \centering
   \centerline{
   \includegraphics[width=0.35\textwidth]{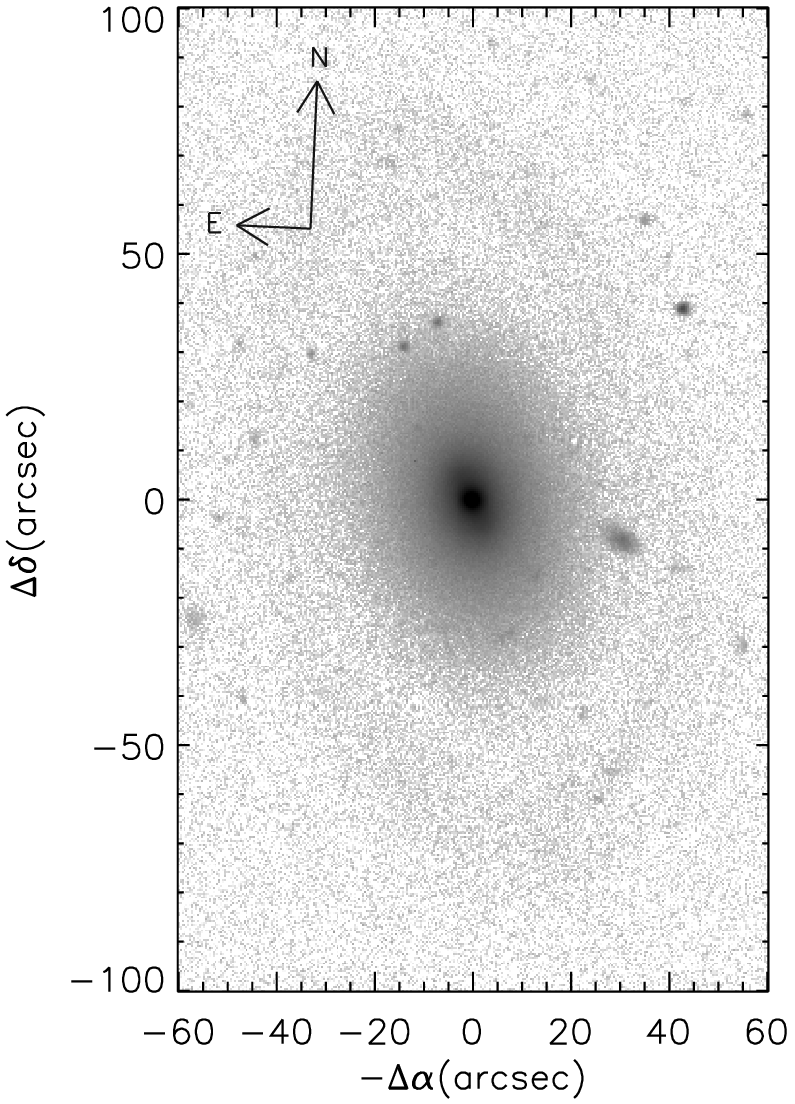}
   \includegraphics[width=0.35\textwidth]{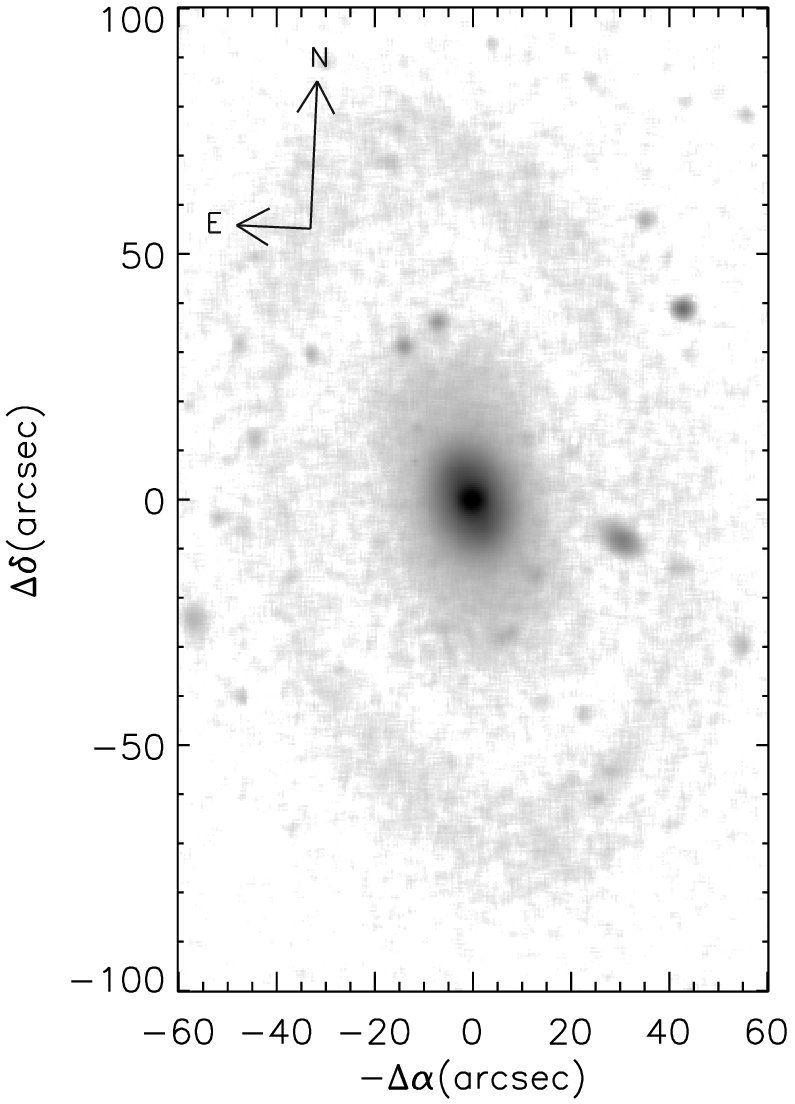}
   }
   \caption{The SDSS $r$-band image of NGC~4513 ({\it the left plot}) and the same image after the subtraction of the outer disc ({\it the right plot}) -- see the Section~3 about the NGC~4513 image decomposition. The brightness scale is logarithmic and the same in both plots.}
 \label{sdss_im}
    \end{figure*}

In this paper we will consider NGC~4513 -- a
northern-sky (R)SA0 galaxy of moderate luminosity, $M_H=-22.8$ (NED). { The image of the galaxy taken from the Sloan Digital Sky Survey (SDSS) DR9 \citep{sdssdr9} is shown in Fig.~\ref{sdss_im}, left, and its main global parameters are given
in Table~1. Earlier we have measured its stellar disc relative thickness
(the ratio of the vertical and radial scalelengths), by using our original method, and have obtained $q=0.245\pm 0.004$ \citep{chud_1}. 
Hence the outer stellar disc of NGC~4513 is somewhat thinner than the bulk of S0 stellar discs in rarefied environments \citep{chud_1,lco_iso} among its morphological 
type \citep{hallsdss}.
The galaxy is rather isolated: according to NED, there is no
galaxies of comparable luminosity within 600~kpc from NGC~4513.} An optical-band ring and the first 
spectral results for this galaxy were reported by \citet{kostuk75} and \citet{kostuk81} long ago.
The galaxy was observed in 21~cm line and was found to be a rather gas-rich S0, with $0.27\times 10^9$\Ms\ of the
neutral hydrogen confined to the ring \citep{tang_hi}.
In the Galaxy Evolution Explorer (GALEX) data, we have detected a UV-ring \citep{we_uvrings} with the radius 
coupled to the optical ring size. Some preliminary description of the spectroscopic results was presented
in \citet{ringmnras}: { it was there that we found that the gas in the outer ring counter-rotated the main stellar body.}. Now we will present a thorough analysis of our long-slit data { and of the structure of the galaxy, including the inner part of NGC~4513} as well as star formation rate (SFR) estimates
{ for the outer ring} obtained from the imaging data in the UV retrieved in the public GALEX archive. The paper is the second one
in the series about star formation in the S0 rings; earlier NGC~6534 and MCG~11-22-015, { two galaxies with corotating detached outer rings,} have been 
described by \citet{s0ring1}. { As we have already found in our previous
consideration of the gas kinematics in the S0 rings \citep{s0_fp}, the gas rotation in the plane coinciding with the stellar disc plane facilitates star formation. Indeed, in NGC~6534 and MCG~11-22-015 we have measured star formation rates of some 0.2~\Ms\ per year that looks rather high for S0 galaxies. As we will show in the present paper, the counterrotating gas of NGC~4513 feeds much weaker star formation.}

\section{Observations and the data involved}

Our long-slit spectral observations were made with a universal
reducer SCORPIO-2 \citep{scorpio2} at the prime focus of the Russian 6-m
BTA telescope of the Special Astrophysical Observatory, Russian
Academy of Sciences. NGC~4513 was observed { on February 8, 2011, putting the $1''$-slit 
along the isophote major axis, $PA(slit)=15\deg$, with the total exposure time of 4800
sec ($4\times 1200$~sec)}.  The seeing during these observations was poor, { $FWHM\approx 3.5$~arcsec}.  
We used the VPHG1200 { grism having the maximum of effectivity at $\lambda \approx 5400$~\AA\ }  providing an intermediate spectral
resolution FWHM $\approx 5$ \AA\ {(corresponding to the instrumental $\sigma$ of 130~\kms)}, to obtain a spectrum in a wavelength region from 4000~\AA\ to 7200~\AA.
This spectral range includes a set of strong absorption and emission lines
making it suitable to analyze both stellar and gaseous
kinematics of the galaxy and its resolved stellar populations.  The slit is $6'$ in
length allowing to use the edge spectra to subtract the sky
background. The CCD E2V CCD42-90, with a format of $2048 \times 4600$ px,
using in the $1\times2$ binning mode provided a spatial scale of 0.357 $''/$px
and a spectral sampling of 0.86~\AA/px.

{
The data were reduced by a standard way using the IDL software package developped in the SAO RAS. At the edges of the slit 
we derived the sky background to subtract it from the  galaxy spectra, by using the polynomial (with
the degree of 4) fit  of the sky background distribution along the slit at every wavelength.
Inhomogeneity of the optics transparency and variations of  the spectral resolution along the 
slit were taken into account by using the dawn spectrum with the high signal-to-noise acquisition. 
The stellar kinematics was calculated by cross--correlating the binned galaxy spectra with the spectra of HD~102328 -- a K2.5-giant star
observed the same night as the galaxy. The emission lines, namely, the H$\alpha$,
[NII]$\lambda$6583, [SII]$\lambda$6717,6731, and [OIII]$\lambda$5007, were used to derive ionized-gas kinematics, by measuring baricenter positions of the lines; in the bins where the continuum is strong we
applied Gauss-analysis to take into account effects of underlying absorption lines: H$\alpha$ as well as
TiI under the [OIII]$\lambda$5007. For the latter purpose, we binned the spectra along the slit to reach signal-to-noise ratio
higher than 50-70, and then made Gauss-analysis of the line complexes:
\begin{itemize}
\item{[NII]$\lambda$6548,6583+H$\alpha$(emission)+H$\alpha$(absorption),}
\item{H$\beta$(emission)+H$\beta$(absorption),}
\item{[OIII]$\lambda$5007(emission)+TiI$\lambda$5007,5015(absorption).}
\end{itemize}
With this analysis we are also able to derive the flux ratios of the strong emission lines: [NII]$\lambda$6583 to H$\alpha$,
[OIII]$\lambda$5007 to H$\beta$, [SII]$\lambda$6717 to [SII]$\lambda$6731,
which have been used to diagnose the gas excitation mechanisms with the BPT-diagrams \citep*{bpt} and to determine electron density and
also gas oxygen abundances for the emission-line regions where the gas is ionized by radiation of young stars.
The detector sensitivity variations along the wavelength were corrected  by observing a spectrophotometric
standard star GRW+70d5824 during the same night. 
}

To study the large-scale structure of the galaxy, we have involved the $g-$ and $r-$band images
from the SDSS DR9 archive \citep{sdssdr9}. To estimate the star
formation rate in the ring, we have retrieved the GALEX
data: NGC~4513 was deeply imaged by this space telescope { on January 21, 2005, in the frame of the Guest Investigator program no. 1-045009 intended to observe another ring galaxy, VII~Zw~466, projected onto the sky plane not far from NGC~4513.
The total exposures of the GALEX observations were}
4286~sec in the FUV-band and 8198~sec in the NUV-band.

\section{The structure of NGC 4513}

\begin{figure*}
\centering
\vspace{0.5cm}
\centerline{ 
   \includegraphics[width=0.33\textwidth]{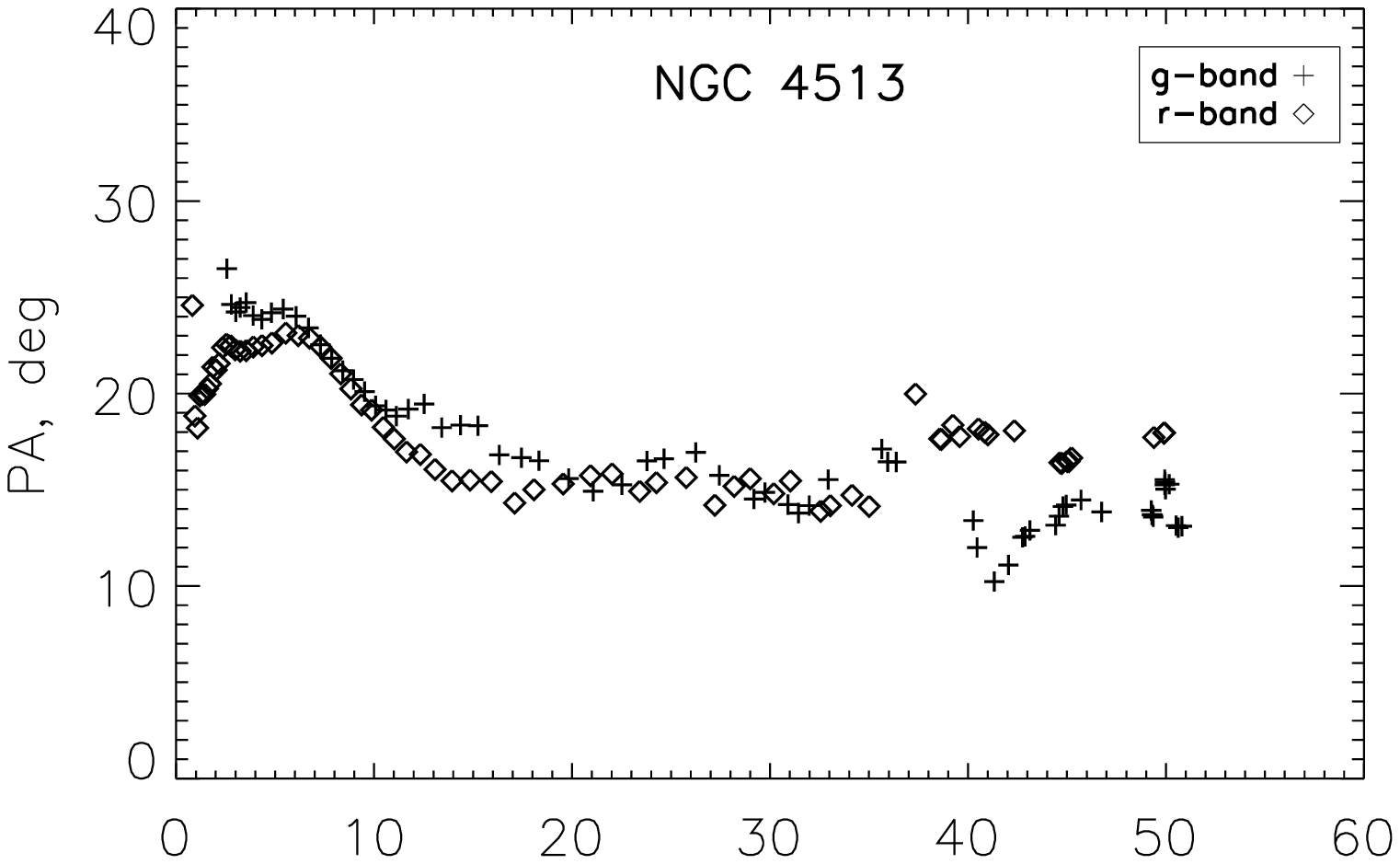} 
   \includegraphics[width=0.33\textwidth]{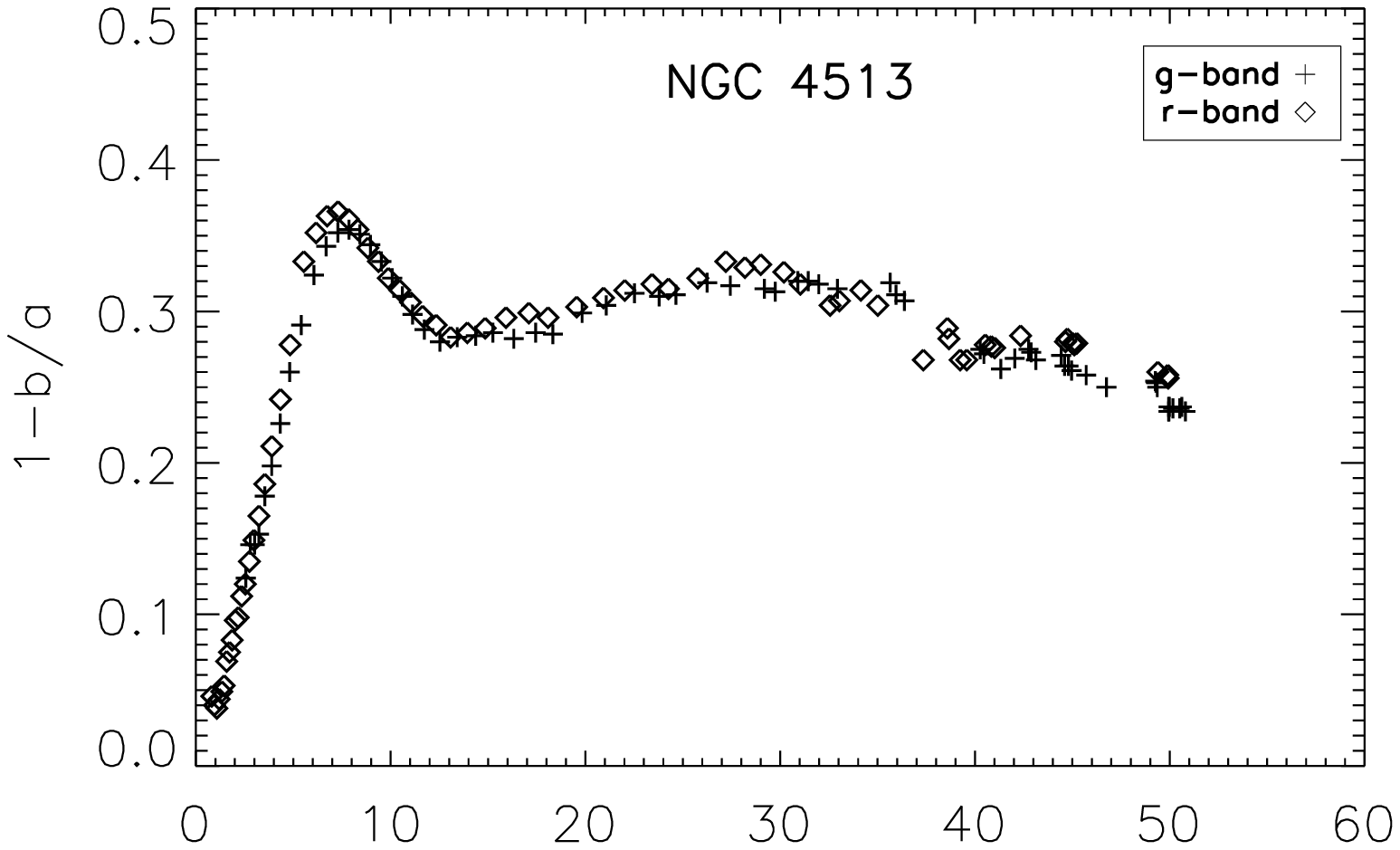}
}
\vspace{0.5cm}
\centerline{
   \includegraphics[width=0.33\textwidth]{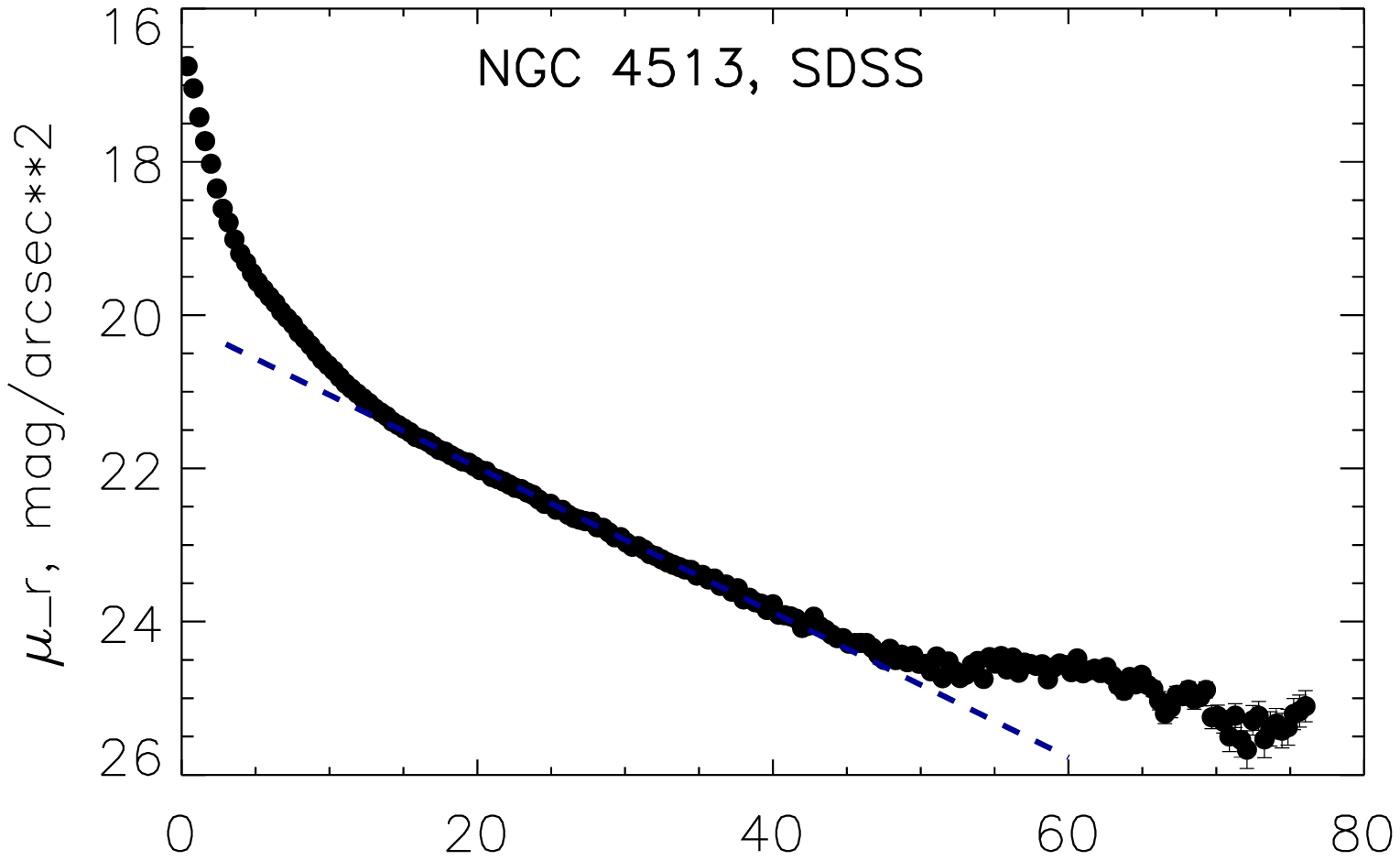} 
   \includegraphics[width=0.33\textwidth]{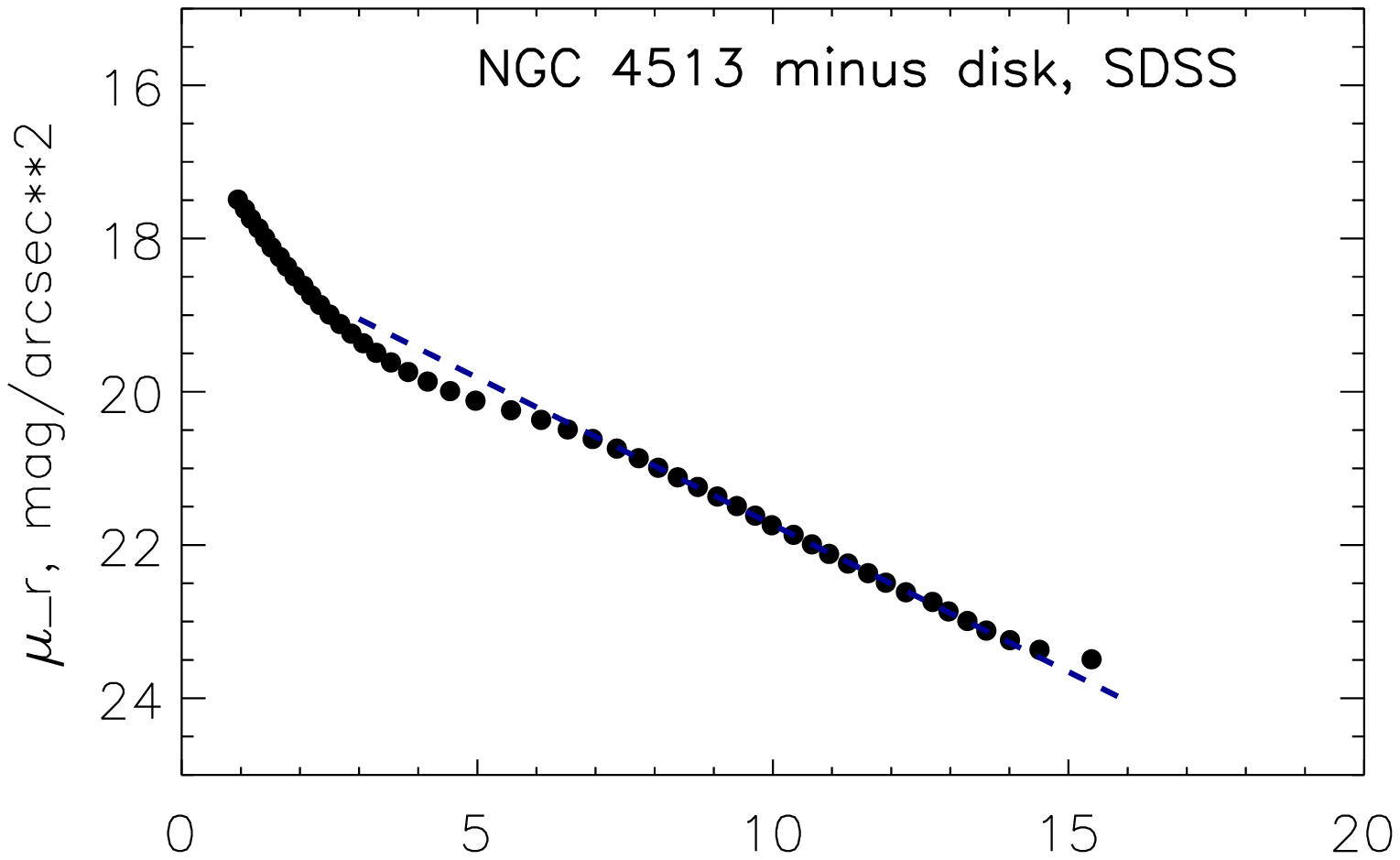}
}
   \caption{The results of the photometric analysis of the SDSS/DR9 data:
   the results of the isophote analysis ({\it upper plots}) and the azimuthally averaged surface-brightness profiles of the full $r$-band image ({\it bottom left plot}) and of the image after the outer disc subtraction ({\it bottom right plot}). The blue dashed lines
   are the fitted
   exponential laws showing the areas of a disc and a pseudobulge domination in NGC~4513.}
              \label{sdss4513}
    \end{figure*}

By using the $g-$ and $r-$band images of NGC~4513 provided by the SDSS/DR9 archive \citep{sdssdr9}, we have undertaken
isophotal analysis, { with the algorithm analogous to ELLIPSE/IRAF}, and then by fixing the isophote parameters, the position angle (PA) of the major
axis and the ellipticity $1-b/a$ of the outermost part { of main body of the galaxy, $R=15\arcsec - 40\arcsec$,} we averaged the surface brightness
over the elliptical rings. { The scatter of the individual ellipticity and position angle measurements around the mean values in the radius range of $R=15\arcsec - 40\arcsec$ allows to estimate the typical errors of $1-b/a$ and $PA$ in the low surface-brightness regions as less than 0.02 and $\sim$ 1\degr, correspondingly. Our logique which we follow during the analysis of galactic exponential discs is presented in detail
by \citet{lcoclust}. The results of the isophote analysis -- the radial profiles of the PA(major axis) and isophote ellipticity, --
as well as the surface brightness profiles} -- are presented in Fig.~\ref{sdss4513}. The local ellipticity maximum and
a turn of the isophote major axis reveal the presence of a bar ending at $R\approx 8$\arcsec.
From $R\ge 14$\arcsec\ to $\sim 40$\arcsec\ the surface brightness profile has a perfect
exponential shape, and the isophote parameters stay constant; we conclude that it is an area
of a large-scale exponential disc domination, { since according to \citet{freeman}, exponential stellar discs are indicated 
by obeying a single-scale exponential law over the radius range of more than twice exponential scalelengths.} The radial scalelength of the exponential profile { of the outer disc of NGC~4513}
is $11\farcs 5$, or 1.8~kpc. { After subtracting the model outer exponential disc
from the complete $r$-band image of NGC~4513, we see a residual image with a rather
diffuse elongated surface brightness distribution (Fig.~\ref{sdss_im}, right). 
By constructing its azimuthally averaged surface brightness profile with ellipse aperture parameters 
running along the radius, we obtain an exponential profile again,
with the scalelength of 2.8 arcsec, or $\sim$0.5 kpc, in the radius
range of 8\arcsec -14\arcsec (Fig.~\ref{sdss4513}, bottom right). We conclude that the bulge of NGC~4513 is in fact a pseudobulge which being a dynamically cold stellar system
includes also a bar.}
In the radius range of $55\arcsec - 72\arcsec$ a surface brightness excess can be noted { in Fig.~\ref{sdss_im} and also in the full surface brightness profile in Fig.~\ref{sdss4513}, bottom left,} -- it is
a signature of an outer stellar ring at the radius of about 10--12~kpc, { beyond the outer edge of the main stellar disc,
$R_{25}=43\arcsec$, or 7~kpc \citep{rc3}.}

\section{Counterrotating gas and complex stellar component in NGC~4513}

\begin{figure}
\centerline{
\includegraphics[width=0.5\textwidth]{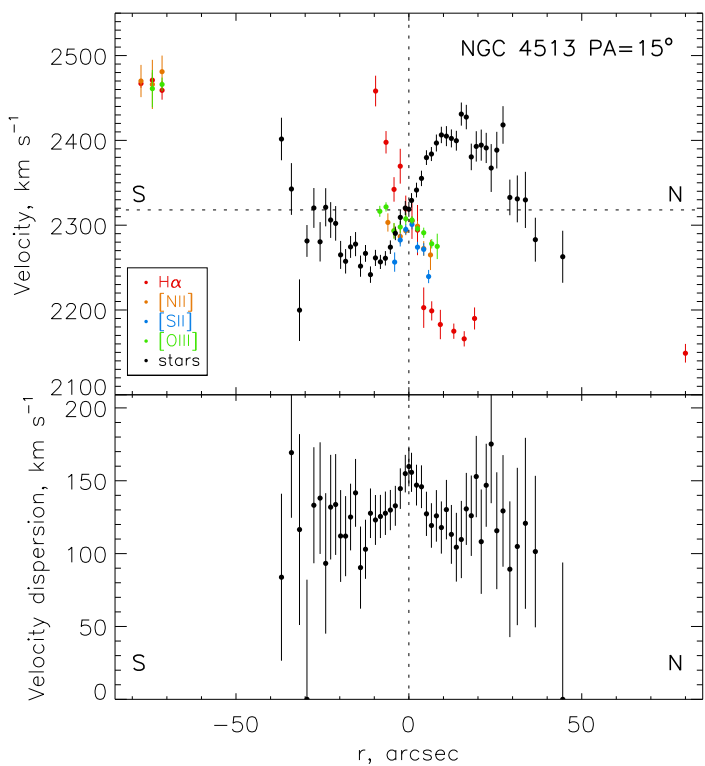}
}
\caption{The line-of-sight velocity  profiles for the ionized gas and stars ({\it top}) and the stellar velocity dispersion profile ({\it bottom}) in NGC~4513
along its major axis. The black signs show the stellar component while various coloured
signs refer to different emission lines of the ionized gas.}
\label{longslit_kin}
\end{figure}

Figure~\ref{longslit_kin} presents the line-of-sight velocity variations along the slit for the stars and ionized gas
in NGC~4513. The velocities of the stellar component were determined by cross-correlating galaxy spectra with the spectrum
of a K-giant star observed the same night with the same spectrograph configuration. The velocities of the ionized gas
in the central part of the galaxy, $R<10$\arcsec, were measured by Gauss-analysis of the line blends including the underlying absorption lines.
In the more outer part { of the main disc, the northern part,} where the continuum is faint we calculated the baricenter wavelength position for H$\alpha$ at every radius.
As Fig.~\ref{longslit_kin} demonstrates, the ionized gas counterrotates the stars over the most extension of the galaxy. The stellar rotation curve { starts to fall} beyond $R\approx 15\arcsec$ and switches into a counterrotating regime at the outer edge of the disc, at $R\ge 30\arcsec$. It implies a possible existence of a secondary
stellar component which may be related to the gas. Its presence may result in superposition of two
counterrotating stellar components giving a null average rotation velocity at $R\approx 25$\arcsec, within the { photometric}
disc-dominated area. { To test this possibility, we have plotted a profile of the measured stellar velocity dispersion estimated as a $\sigma$ of the stellar LOSVD (Fig.~\ref{longslit_kin}, {\it bottom}). 
Though our spectral resolution does not allow to measure reliably stellar velocity dispersions below 100~\kms\ expected in a disc, we can nevertheless feel qualitatively some increase of the visible stellar velocity dispersion after the reverse of the rotation curve.
Such behaviour of the stellar velocity dispersion profile
supports our suggestion about two stellar-rotation components at our line of sight in the disc-dominated area.
The rotation of the ionized gas in NGC~4513 is traced by measuring
four strong emission lines (Fig.~\ref{longslit_kin}). In the central
part of the galaxy, $R < 10\arcsec$, we see a rather flat segment of the gaseous velocity profile,
consistent with a suggested gas slowdown at the bar edges.
Further in the pseudobulge area the gas rotation curve rises steeply, and} 
the gas velocities in the ring, { consistent with the gas velocities at $R\ge 10\arcsec$,} give evidence for a intrinsically flat character of the rotation curve over the
full extension of the NGC~4513 disc. { We do not see any emission lines between the inner edge of the disc and the outer ring;
it is in line with the finding by \citet{tang_hi} on a prominent
central depression in HI gas disc.}

\section{Stellar population properties}

\begin{figure}[htb!]
\vspace{0.5cm}
   \centering
   \begin{tabular}{cc}
   \includegraphics[width=0.22\textwidth]{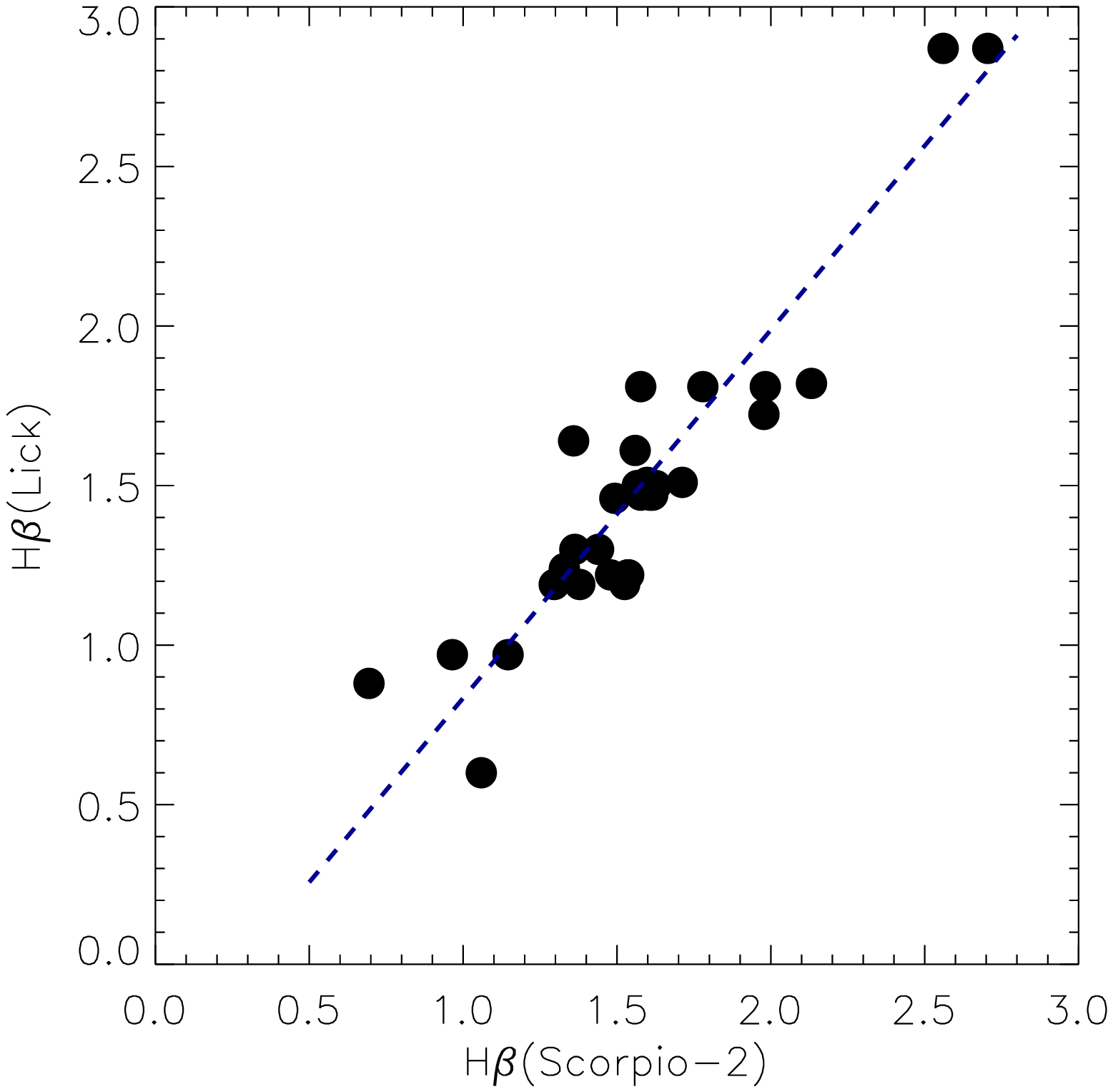} &
   \includegraphics[width=0.22\textwidth]{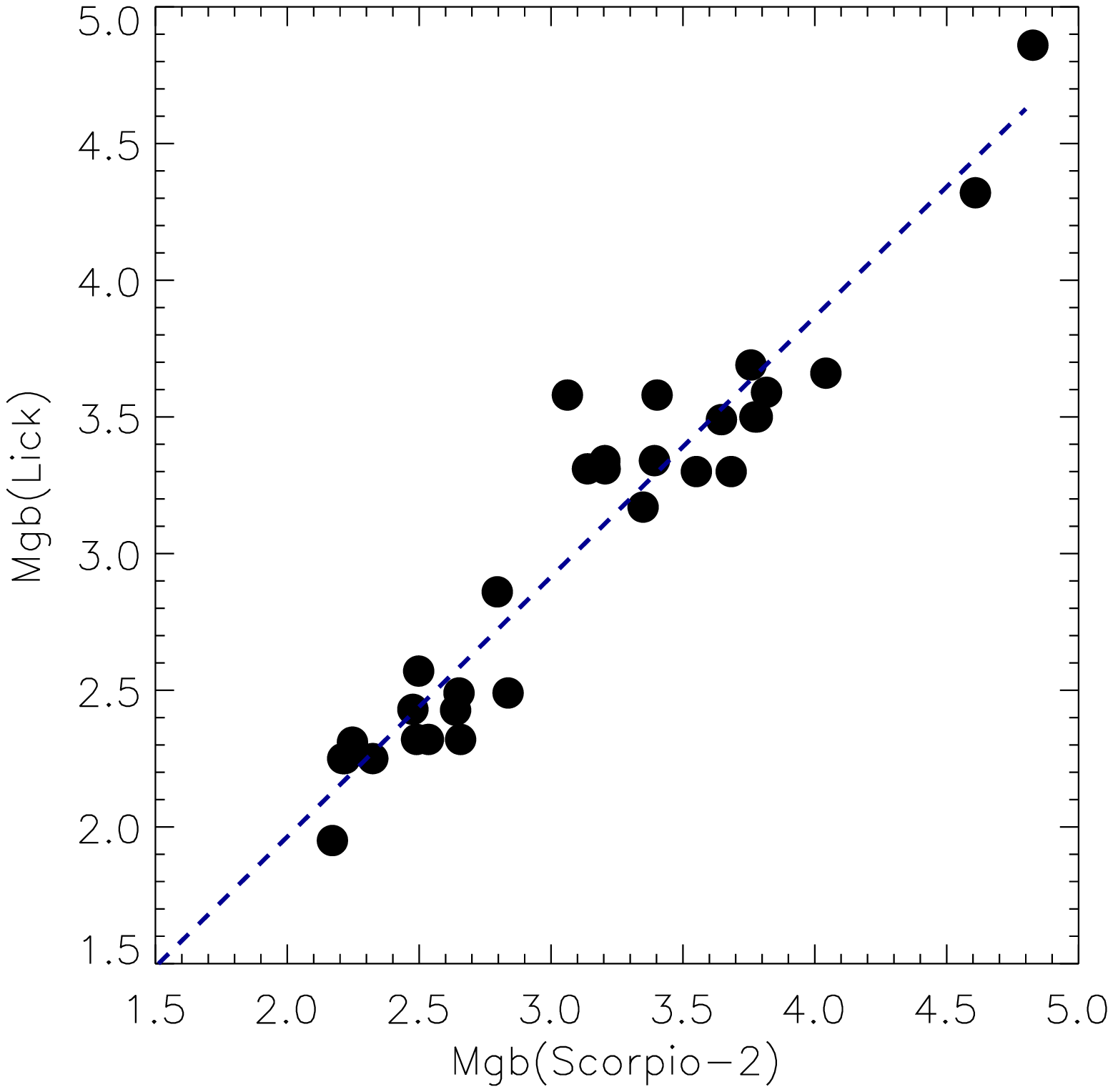}\\
    & \\
   \includegraphics[width=0.22\textwidth]{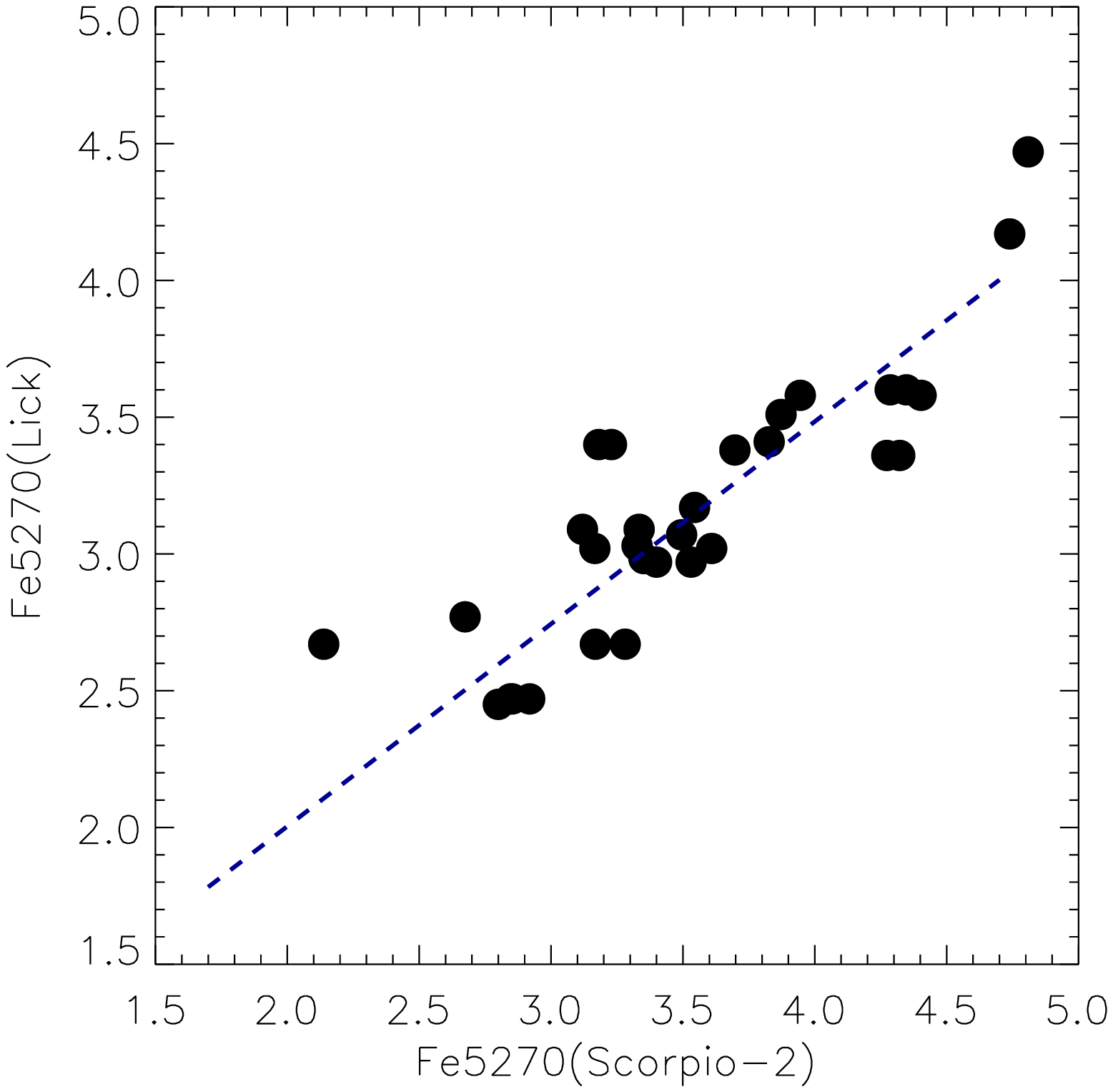} &
   \includegraphics[width=0.22\textwidth]{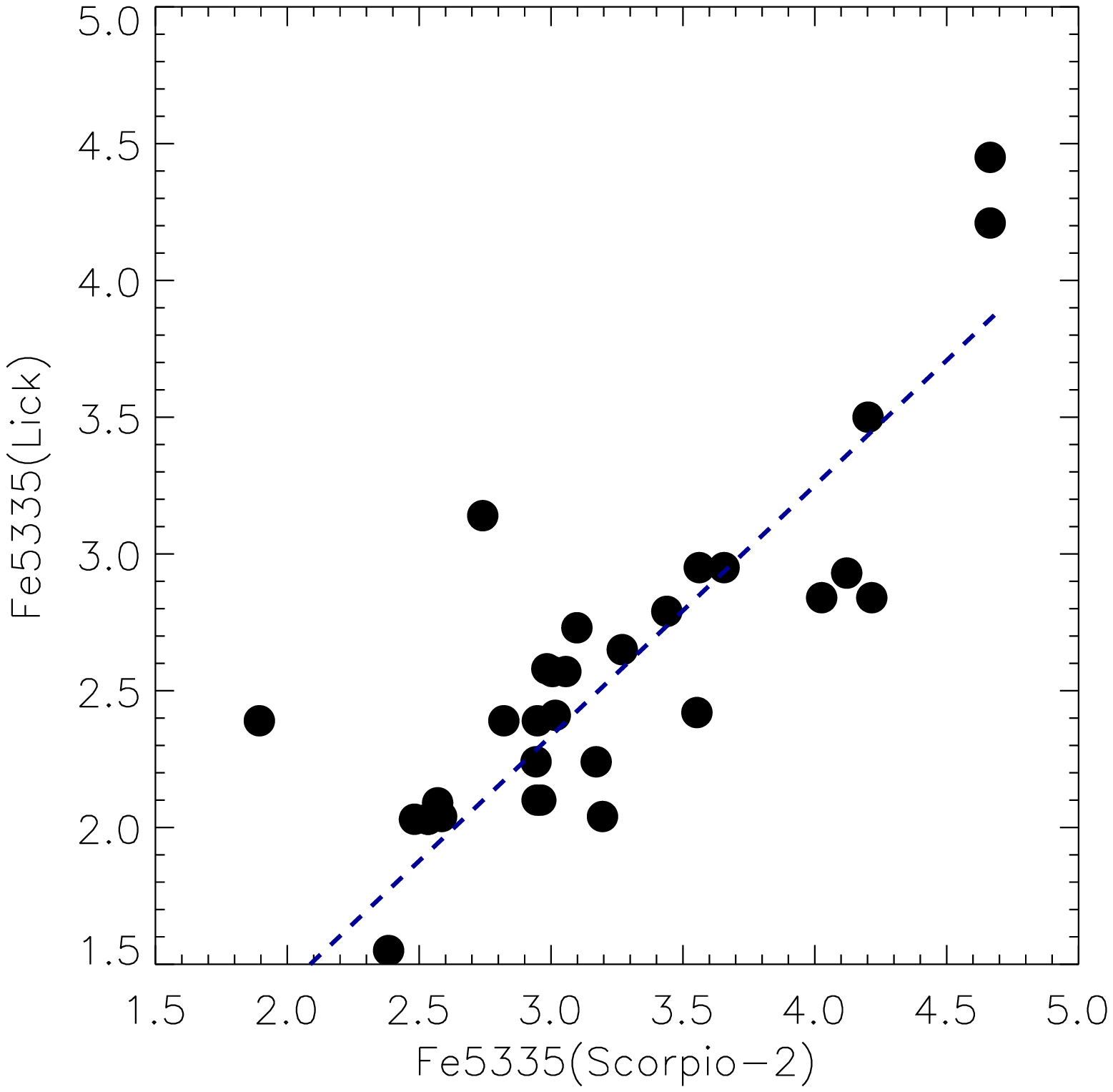}\\
   \end{tabular}
   \caption{The linear calibration of our instrumental Lick indices
   versus standard Lick indices for a sample of bright stars
   from \citet{woretal}.}
              \label{indsys}
    \end{figure}

{ To analyze the ages and chemical composition of the NGC~4513
stellar population, we have calculated the Lick indices H$\beta$,
Mgb, Fe5270, and Fe5335, as well as the combined iron index,
$\langle \mbox{Fe} \rangle \equiv (\mbox{Fe5270} + \mbox{Fe5335})/2$, and combined metallicity index,
[MgFe]$\equiv \sqrt{\mbox{Mgb} \langle \mbox{Fe} \rangle}$,
along the radius by using our long-slit
spectrum obtained with the SCORPIO-2 of the Russian 6m telescope.
We prefer here the Lick index analysis because we expect non-solar
ratio of the $\alpha$-element-to-iron abundances, and popular now
full-spectral fitting method is still restricted to the solar element pattern assumption. Firstly, we have calibrated our SCORPIO-2 index
system to the standard Lick one by observing a sample of standard
stars from the \citet{woretal}. The linear calibration dependencies are presented in Fig.~\ref{indsys}. The scatter of individual stars around the linear dependencies, $\sim 0.2$~\AA, is comparable to the accuracy of the Lick index measurements by \citet{woretal}.}

\begin{figure*}[htb!]
   \centering
   \centerline{
   \includegraphics[width=0.35\textwidth]{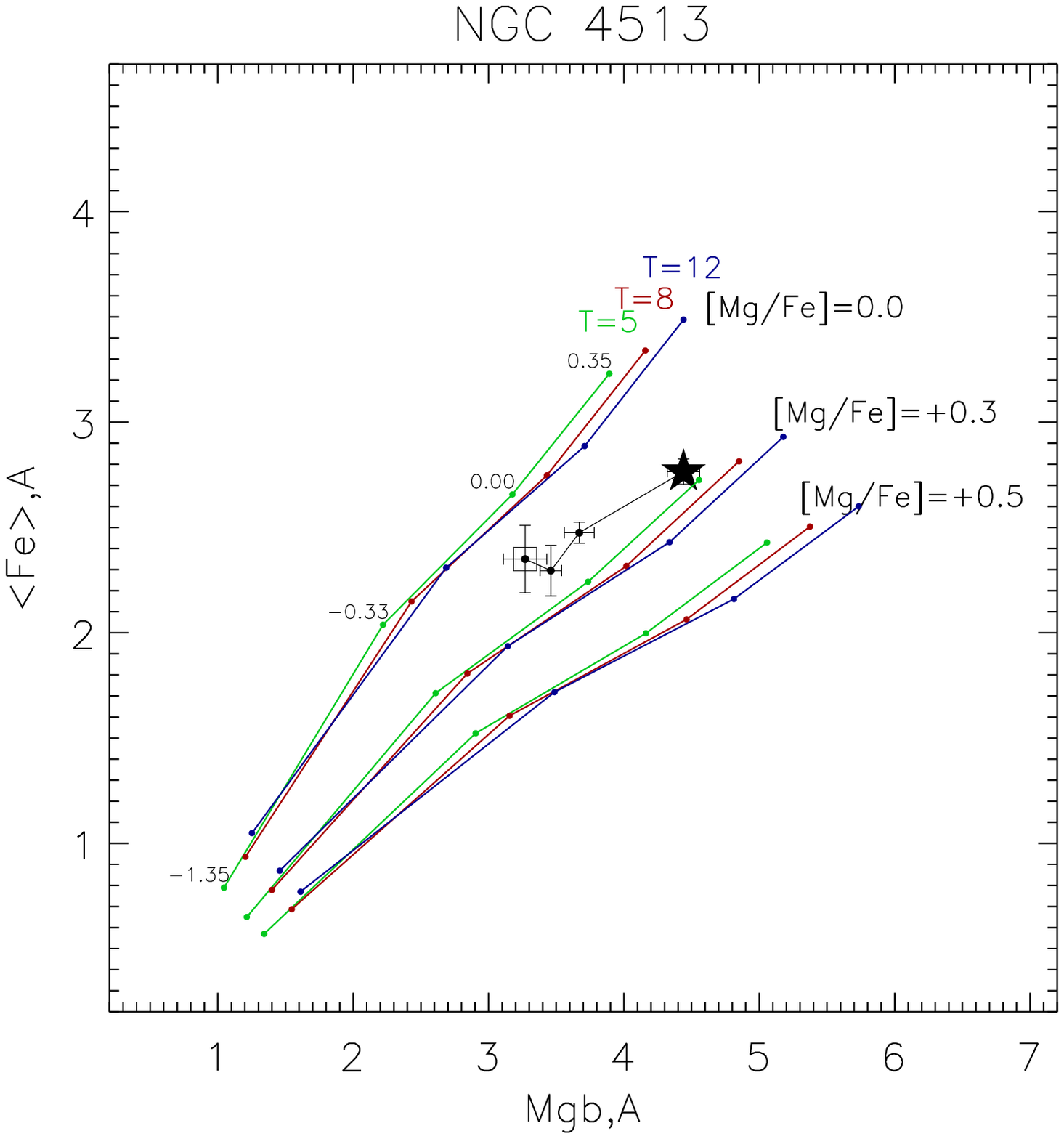}
   \includegraphics[width=0.35\textwidth]{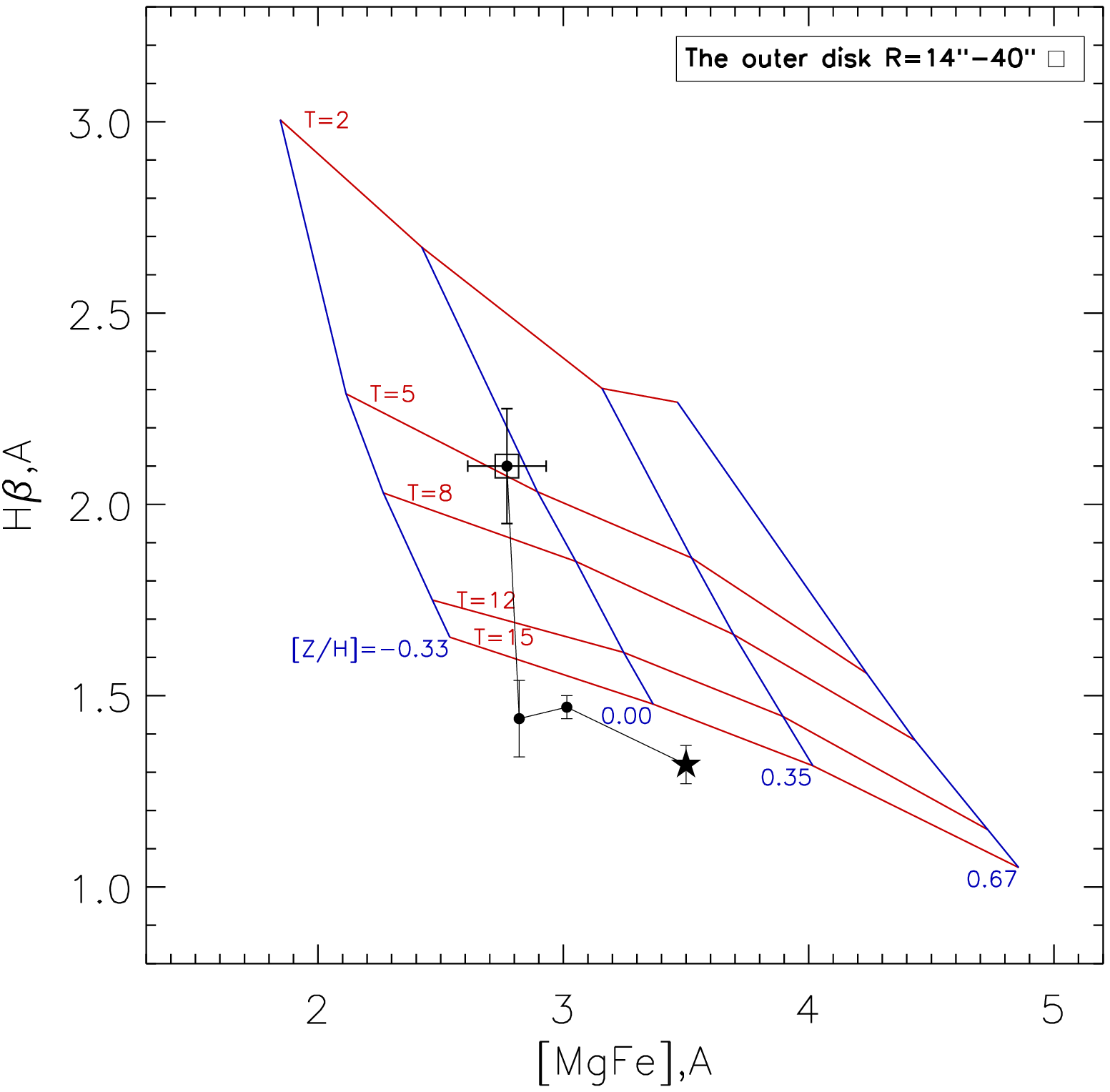}
   }
   \caption{Lick index-index diagrams for NGC 4513. The {\it left plot} represents Mgb vs iron index diagram which allows to estimate magnesium-to-iron ratio through the comparison of our measurements with the models by \citet{thomod} for the different Mg/Fe ratios. By confronting the H$\beta$ Lick index versus
a combined metallicity Lick index involving magnesium and iron lines ({\it right plot}), we solve the metallicity-age degeneracy
and determine these stellar-population parameters with the SSP evolutionary synthesis models by \citet{thomod}.
Five different age sequences (red lines) are plotted as reference frame; the blue lines crossing the model age
sequences mark the metallicities of $+0.67$, $+0.35$, 0.00, --0.33 from right to left. A large black star corresponds
to the central core, $R<4$\arcsec, and then we go along the radius through the galaxy structure components:
$R=4^{\prime \prime}-8$\arcsec (bar), $R=8^{\prime \prime}-14$\arcsec (pseudobulge), and $R=14^{\prime \prime}-40$\arcsec (disc). The point corresponding to the outer
stellar disc is outlined by a large square.}
              \label{lickind}
    \end{figure*}

Then we have estimated the SSP-equivalent ages and metallicities of the stellar populations { along the radius} of NGC~4513
by confronting our measurements of the Lick indices to the evolutionary synthesis models by \citet{thomod}. 
{ The H$\beta$ index was corrected for the emission contamination in the innermost part of the galaxy through measuring the H$\alpha$ emission-line equivalent width as it was described in \citet{me2006}.
After that, broad radial bins corresponding to the photometric borders of the unresolved nucleus, $R<4$\arcsec, the bar, $R=4^{\prime \prime}-8$\arcsec, 
the pseudobulge, $R=8^{\prime \prime}-14$\arcsec, and the large-scale stellar disc, $R=14^{\prime \prime}-40$\arcsec,
were defined. Our Lick indices were averaged within these radial
bins.} The results are
presented in Fig.~\ref{lickind}. The central part of the galaxy is homogeneously old and magnesium-overabundant. However, the
large-scale stellar disc of NGC~4513 differs from the center of the galaxy as concerning the properties of the stellar
population: it reveals a more prolonged history of star formation (its magnesium-to-iron ratio is closer to the solar value),
which had stopped only a few billion years ago. The stellar metallicity of the galactic disc is only slightly subsolar.
{ It is rather unusual for a lenticular galaxy in rarefied
environments where the outer stellar discs are found to be coeval with the bulges \citep{isosalt} or are older than the bulges \citep{s0disc12}. We may relate this unusual stellar age distribution again with the presence of the secondary, probably young stellar component which has come into the outer disc of NGC~4513 with the counterrotating gas accretion.}

\section{Gas-phase metallicity}

\begin{figure}
   \centering
   \centerline{
   \includegraphics[width=0.2\textwidth]{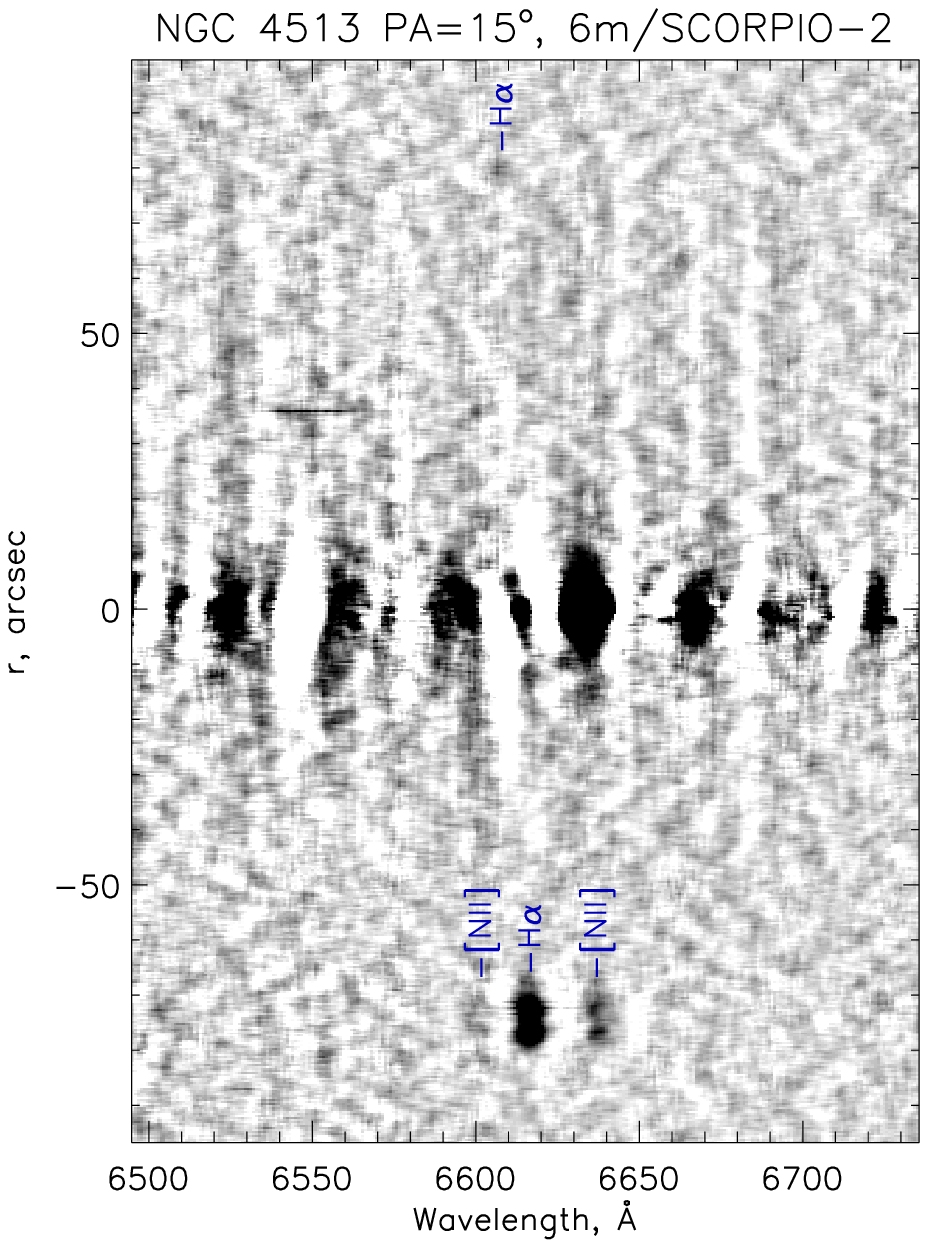}
   \includegraphics[width=0.25\textwidth]{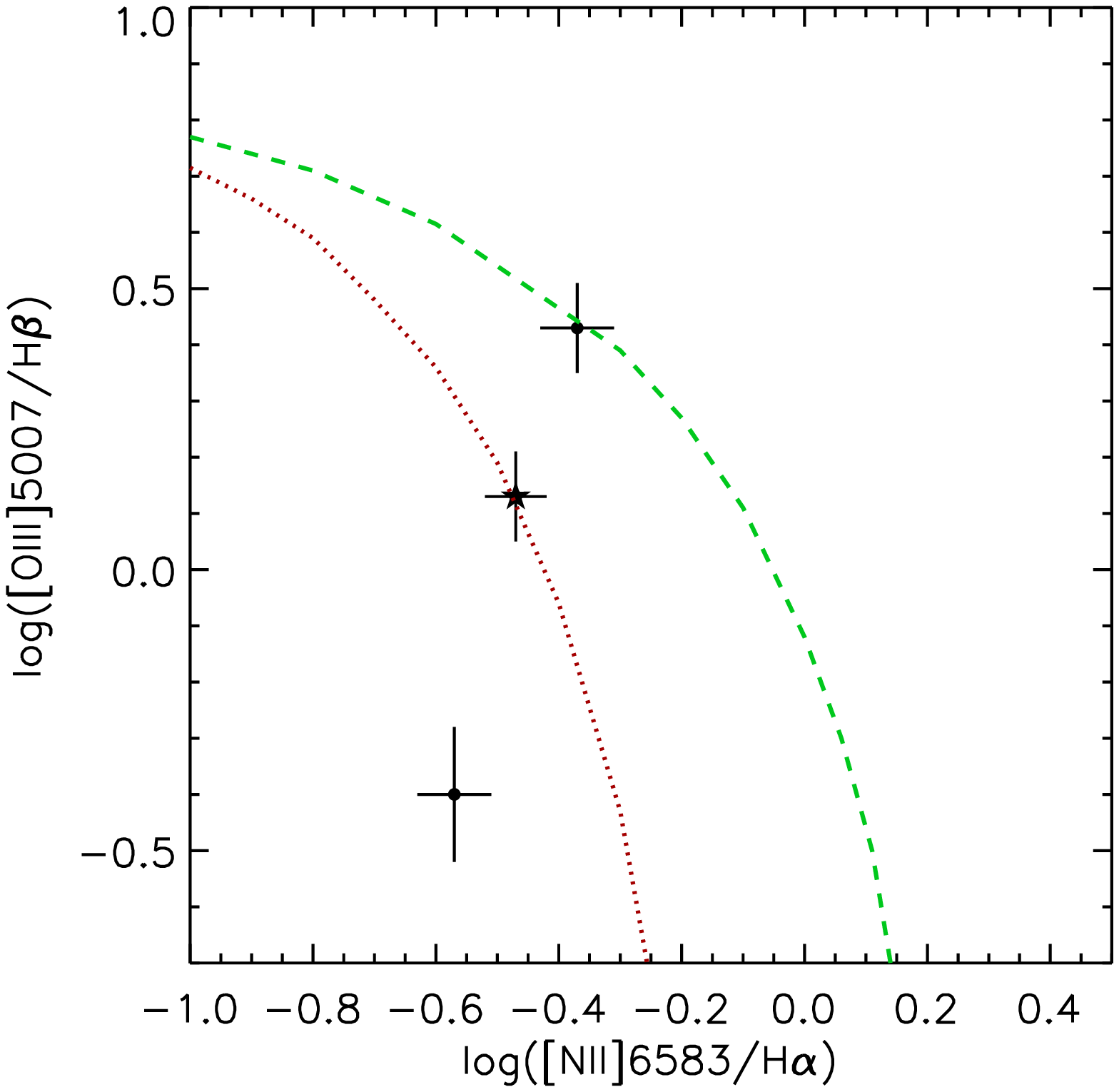}
   }
   \caption{Emission lines in the spectrum of NGC 4513: H$\alpha$ and the [NII] doublet ({\it left})
and the BPT-diagram for the southern tip of the ring ({\it right}.) At the BPT-diagram the known boundaries
between HII-region excitation type and the others are plotted: the green dashed line is a theoretical
boundary by \citet{kewley01}, and the dotted red line is an empirical star-formation sequence from
\citet{kauffmann03}. We have plotted separately the inner edge of the ring,
$R=71.5\arcsec$ {\it(upper dot)}, and the outer edge of the ring, $R=76.5\arcsec$ {\it(lower dot)}. The whole (integrated) ring is plotted by a black star.
}
 \label{ringbpt}
 \end{figure}

{ Figure~\ref{ringbpt}, {\it left}, shows the emission-line long-slit spectrum of NGC~4513 along its major axis, namely, its red portion with the continuum subtracted; and Fig.~\ref{ringbpt}, {\it right}, -- the BPT-diagram for the lowest (southern) location of the emission lines. One can clearly see gas-star counterrotation as well as the absence of the emission lines between the central part of NGC~4513 and its starforming ring which demonstrates emission lines at the $R=70\arcsec - 80\arcsec$.}

In the central part of NGC~4513 the emission line \Ha\ is everywhere weak, and the strongest emission line is [NII]$\lambda$6583.
Such line ratio is consistent with the possible gas excitation by old stars \citep{pagb94,liermodel} that is in agreement 
with the age of the stellar population in the central part of NGC~4513 (see Section~5). Otherwise the gas within $R<8$\arcsec\
could be excited by shock mechanism which is consistent with the bar presence. The bar contribution into the gas excitation is 
probably also manifested by strongly asymmetric electron density distribution along the slit: we have measured the trend of 
the sulfur line ratio, [SII]$6717/6731$, from $0.71\pm 0.07$ to the south from the nucleus to $1.13\pm 0.04$ to the north 
from the nucleus that corresponds to the $n_e$ difference of about an order -- 2400~cm$^{-3}$ versus 300~cm$^{-3}$ \citep{kewley19}.

In the ring the situation with a source of gas ionization could be different -- here we expected the gas excitation by current
star formation. We have plotted the emission-line ratios in the southern tip of the ring onto the BPT diagram (Fig.~\ref{ringbpt}, {\it right}). Indeed, the strong-line ratios in the ring of NGC~4513 have appeared to lie below the theoretical border
of the star formation calculated by \citet{kewley01} so the ionized gas of the ring may be mostly excited by young stars. But the 
{ prominent} offset of the
inner edge of the ring at the BPT-diagram with respect to the observational star formation sequence by \citet{kauffmann03} puts this region
into the so called 'composite zone' revealing a noticeable contribution of diffuse interstellar gas (DIG) or shocks into the spectrum of the ionized gas of the ring. It means that not all strong-line methods of the gas oxygen determination are applicable to the inner edge of the ring in NGC~4513.
Recent studies have shown that in the presence of DIG the most safe metallicity calibration is provided by the O3N2 method
\citep{kumari}, and we have decided to use just this method. By exploring both O3N2 calibrations from \citet{pp04} and \citet{marino13},
we have obtained for the inner edge of the ring of NGC~4513 $12+\log \mbox{(O/H)} =8.42\pm 0.06$~dex. For the outer edge of the ring which is completely in the HII-region area of the BPT-diagram and for the whole ring which falls exactly onto the star formation sequence by \citet{kauffmann03} we have used both N2 and O3N2 methods from the papers by \citet{pp04} and \citet{marino13}. We have obtained $12+\log \mbox{(O/H)} =8.57\pm 0.06$~dex for the ring outer edge and $12+\log \mbox{(O/H)} =8.54\pm 0.06$~dex for the whole ring. { Together with the inner edge of the ring, these estimates imply
the ring gas metallicity of} about --0.2~dex with respect to the solar metallicity.
It matches rather closely the stellar metallicity of the large-scale disc of NGC~4513 as reported in the Section~5. { If to compare
the gas oxygen abundance in NGC~4513 to the other outer rings of S0s studied by us \citep{wesaltrings,s0_fp}, it is just the same metallicity: for a sample of dozen gaseous outer rings in S0s, mostly with
corotating kinematics (though in NGC~2551 the gas counterrotates), we have found $\langle \mbox{[O/H]} \rangle =-0.15$~dex \citep{s0_fp}.}

 \section{Star formation in the outer ring of NGC 4513}

\begin{figure*}
   \centering
   \centerline{
   \includegraphics[height=0.4\textwidth]{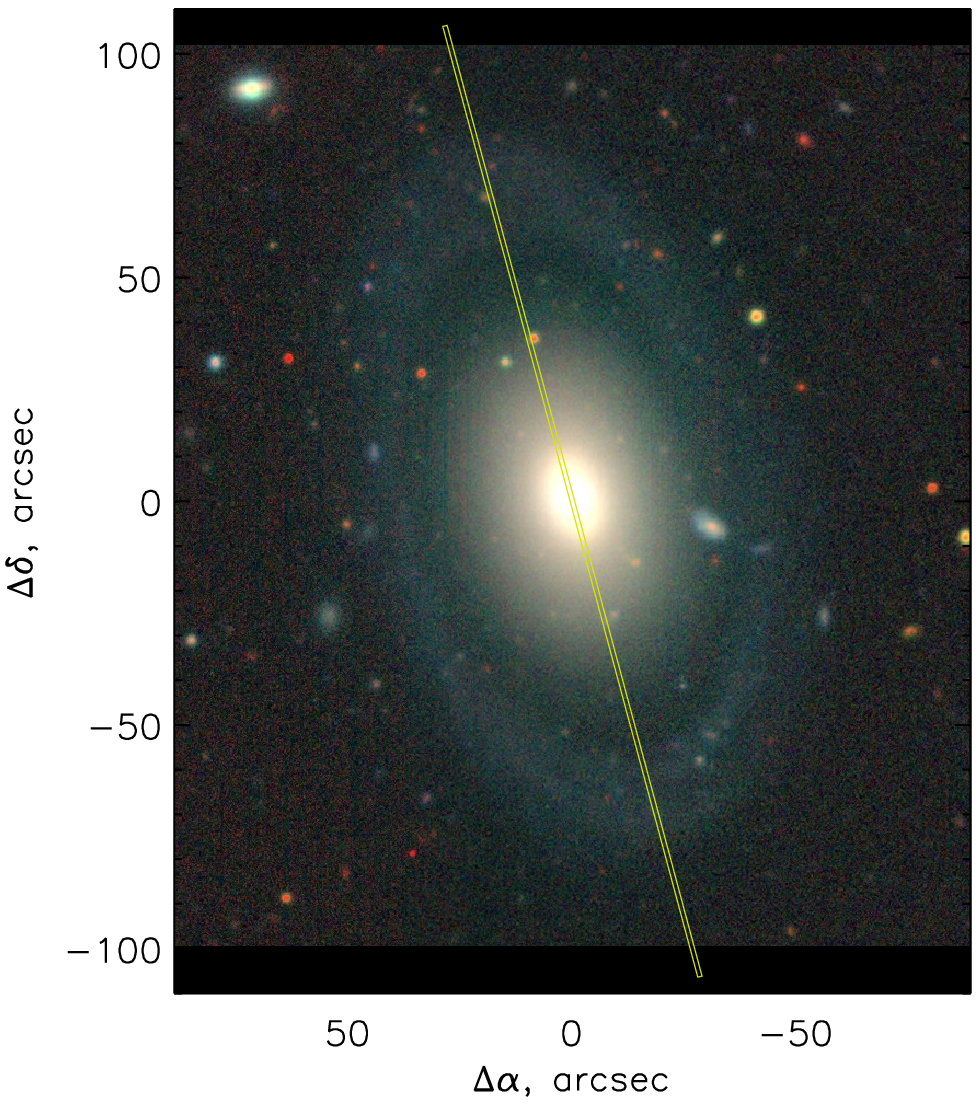}
   \includegraphics[height=0.4\textwidth]{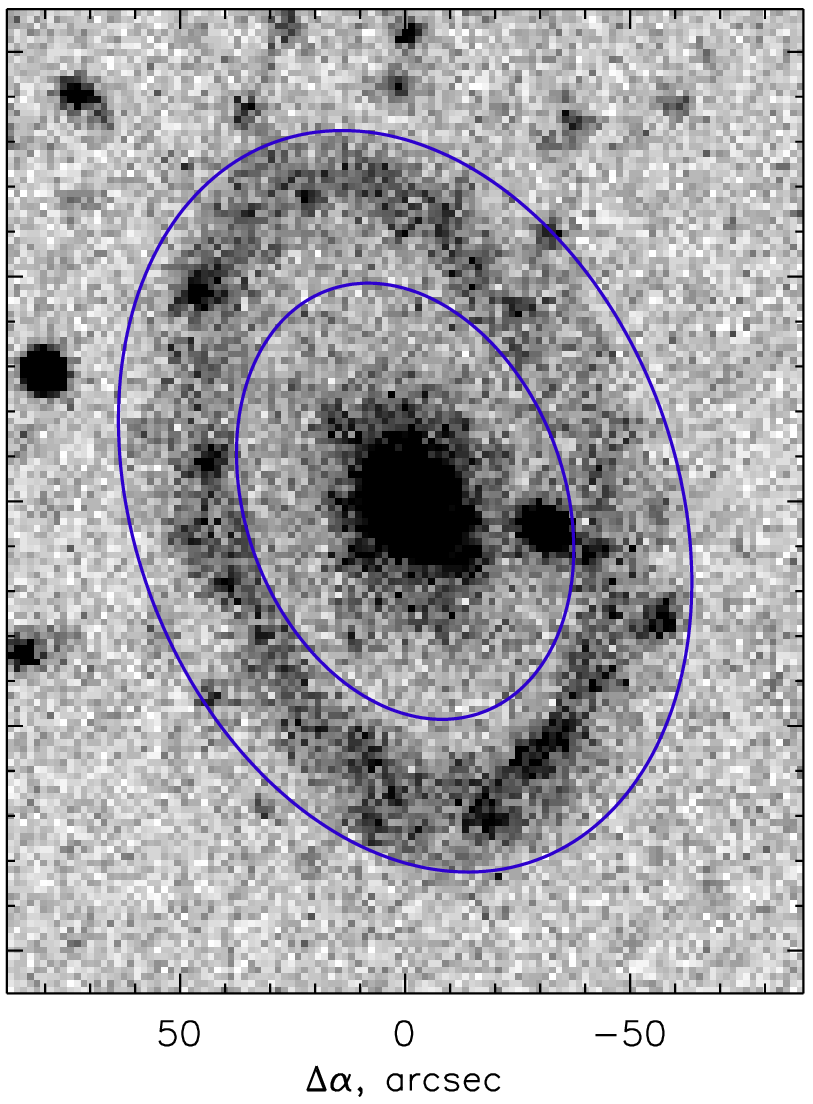}
   \includegraphics[height=0.4\textwidth]{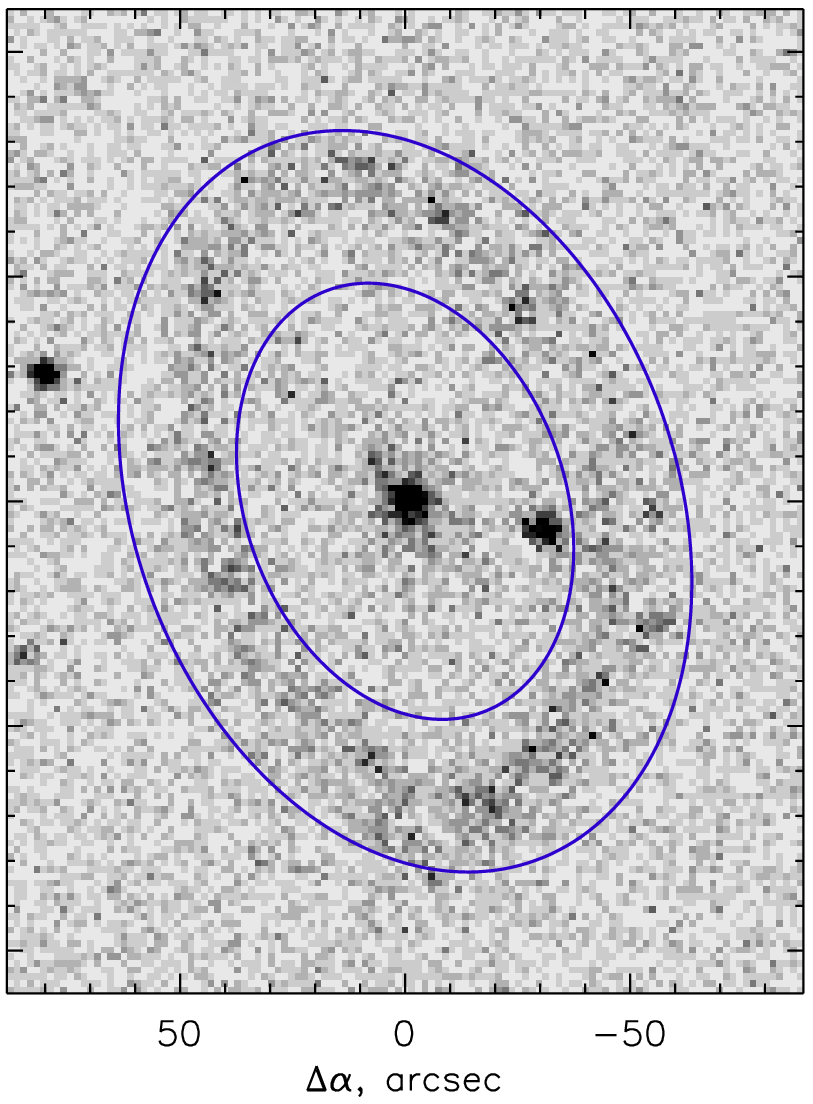}
   }
   \caption{The optical composite-color image ({\it left}), the product of the BASS survey, is taken from the Legacy 
Survey website (http://legacysurvey.org), the SCORPIO-2 slit position is overlapped; as well the GALEX maps of NGC~4513 are shown with the overposed apertures 
for the SFR estimates in the ring, NUV ({\it middle}) and FUV ({\it right}).
}
 \label{galex_im}
 \end{figure*}

NGC~4513 has been observed by the UV space telescope GALEX with rather large exposure times, so both the FUV- and NUV-images of the galaxy are deep, and the UV-appearance of the outer ring of NGC~4513
is rather prominent (Fig.~\ref{galex_im}). We have retrieved these FUV- and NUV-image from the MAST Archive and
have overposed onto them an elliptical-ring aperture, { slightly
broader than the ring itself, to include all the ultraviolet flux}, with the ellipticity matching the galaxy inclination, $1-b/a=0.3$.
The aperture is centered onto the NGC~4513 nucleus, aligned with the outer isophote major axis, have the inner radius of 50\arcsec\
and the outer radius of 85\arcsec\ (Fig.~\ref{galex_im}). And then we have integrated the FUV- and NUV-fluxes
within this ring aperture. The surrounding background was measured and subtracted. The fluxes in counts have been re-calculated
into FUV- and NUV-magnitudes using the procedures described by \citet{galex}. Then we have corrected them for the foreground
Galactic extinction by taking the NED $A_B$ data for NGC~4513, and transformed into FUV- and NUV-luminosities by using
the NED-provided distance of 33~Mpc to NGC~4513. We have applied the correction for the intrinsic dust by using
the WISE/Band 4 (22$\mu$m) image of NGC~4513 cutted with the same elliptic-ring aperture. By obtaining 
the FUV- and NUV-luminosities of the ring, we have transformed them into the star formation rates averaged over the last 100
and 200~Myr respectively, by using the calibrations proposed by \citet{ken_evans}.

The resulting SFR estimate for the ring of NGC~4513 is { 0.026} solar mass per year (0.022\SFR\ from the FUV-data and
0.030\SFR\ from the NUV-data). We have compared this SFR with the total stellar mass of the NGC~4513 ring. Indeed,
the SDSS data allow to estimate the integrated $g$-band and $r$-band magnitudes of the ring (in the same elliptical-ring
aperture as the UV-signals). They have appeared to be $g(ring)=15.16$ and $r(ring)=14.61$. With the distance to NGC~4513
of 33~Mpc, we derive the absolute magnitude of the ring $M_g(ring)=-17.4$. By using the \citet{bell03} calibration
of the mass-to-luminosity ratio against the colour, having $g-r(ring)=0.55$, we assume $M/L_g=2.14$. Then the total
stellar mass of the ring is $2.26 \cdot 10^9$\Ms. 

{ Now we can try different scenarios of the star formation
history in the ring.} 
To accumulate { the stellar mass of $2.26 \cdot 10^9$\Ms }  with the constant SFR in the ring of 0.026\SFR { -- the rate which we have found from the UV-signal for the last 100-200~Myr, --}
the galaxy needs much more than the Hubble time. 
{ Under the opposite scenario}, if the star formation history (SFH) declined exponentially
and started some 3~Gyr ago, we would obtain the same stellar
mass with a e-folding time of 0.6~Gyr which is rather typical for the S0 outer ring SFH \citep{wesaltrings}.  In the former
scenario we must conclude that not only the gas has been accreted from outside -- a substantial amount of the stellar
mass has been accreted too. In this case we deal not with pure gas accretion, but with tidal disruption
of a gas-rich satellite. In the latter scenario, the galaxy can accrete only cold gas, but the mass of this gas,
about $2.5\times 10^9$\Ms, constitutes more than 10\%\ of the total stellar mass of NGC~4513. The invisible source of the counterrotating
gas in this case must be very abundant. In principle, it can be accretion of primeval gas from a cosmological filament
because the current ratio of stellar-to-gas mass in the ring, $\sim 8$, puts the chemical evolution of the ring of NGC~4513 into
the 'gas-depleted' stage \citep{zahid14,ascasibar15} so the current metallicity of the ionized gas must correspond to the
saturation level which is close to the solar value \citep{zahid14,ascasibar15}, independently of the initial gas metallicity.
It may explain the proximity of the gas oxygen abundance in the ring of NGC~4513 to a common value found by us for a number
of other outer rings in S0 galaxies \citep{s0_fp}.

\section{Discussion and conclusion: The origin of the ring in NGC 4513}

Though according to our photometric analysis results (Fig.~\ref{sdss4513}) NGC~4513 has a small bar, however we do not think that
its outer ring relates somehow to the bar's resonances. Indeed, rings at outer Lindblad resonances show typical radii of 1.5--2
bar radius \citep{buta17}, while in NGC~4513 the ratio { $R(ring)/R(bar) =10\pm 2$.} 
Furthermore, the gas content of the ring counterrotates the main stellar body of the galaxy; this fact gives unambiguous evidence 
for an accretion nature of the ring.

The star formation in the ring is weak, 0.026\SFR over a timescale of some 100~Myr.
If such level of SFR remained roughly constant, the stellar content of the ring, some $2.3 \cdot 10^9$\Ms,
could not be formed {\it in situ} and had to be also accreted together with the gas. If the SFH in the ring was strongly
declined, with a e-folding time of $\sim 0.6$~Gyr during the last 3~Gyr, the stellar content of the ring could be formed {\it in situ}.
The latter scenario implies that the stellar ring of NGC~4513 may be a consequence of a long star-formation event
provoked perhaps by gas accretion from a cosmological filament as in the ring-like Hoag galaxy \citep{hoag}.
In any case, the stars related to the counterrotating gas may contribute to the outer stellar disk of NGC~4513 
which demonstrates the SSP-equivalent stellar ages 
of less than 5~Gyr together with the sharply falling rotation curve. The most probable scenario of the NGC~4513 ring 
acquisition is a tidal disruption of a gas-rich satellite. The ratio $\log M_{HI}/M_* \ge -0.9$ which we have found
for the outer ring of NGC~4513 is indeed quite typical for satellite galaxies in gas-rich loose groups \citep{dzudzar}. 
The environment of NGC~4513 corresponds to this scenario: according to the NED, the galaxy is rather isolated, it
belongs to a loose group of 4 galaxy, and the nearest neighbour, emission-line dwarf PGC~2683704, is in 270~kpc and less massive 
by a factor of 30. Perhaps, a similar but a more close satellite was disrupted by NGC~4513 and had formed a ring with the radius 
corresponding to its orbital momentum.

\begin{acknowledgements}
{ We are grateful to the anonymous referee who has made a lot of
very useful comments resulting in the paper improvement.}
The study of galactic rings was supported by the Russian Foundation for Basic
Researches, grant no. 18-02-00094a. The work is based on the data obtained at the Russian
6m telescope of the Special Astrophysical Observatory carried out with the financial support of the Ministry of Science and Higher Education of the Russian Federation and on the public data
of the SDSS (http://www.sdss3.org) and GALEX (http://galex.stsci.edu/GR6/) surveys.
The NASA GALEX mission data were taken from the Mikulski Archive for Space
Telescopes (MAST). The WISE data exploited by us were retrieved from the NASA/IPAC Infrared Science Archive,
which is operated by the Jet Propulsion Laboratory, California Institute of Technology,
under contract with the National Aeronautics and Space Administration. The NGC~4513 composite-colour optical image was taken
from the Legacy Survey collection providing the imaging data of the BASS survey.
BASS is a key project of the Telescope Access Program (TAP), which has been funded by
the National Astronomical Observatories of China, the Chinese Academy of Sciences
(the Strategic Priority Research Program "The Emergence of Cosmological Structures" Grant no. XDB09000000),
and the Special Fund for Astronomy from the Ministry of Finance. The BASS is also supported
by the External Cooperation Program of Chinese Academy of Sciences (Grant no. 114A11KYSB20160057),
and Chinese National Natural Science Foundation (Grant no. 11433005).
The Legacy Surveys imaging of the DESI footprint is supported by the Director, Office of Science,
Office of High Energy Physics of the U.S. Department of Energy under Contract No. DE-AC02-05CH1123, 
by the National Energy Research Scientific Computing Center, a DOE Office of Science User Facility 
under the same contract; and by the U.S. National Science Foundation, Division of Astronomical Sciences
under Contract No. AST-0950945 to NOAO.
\end{acknowledgements}

%
%

\end{document}